\documentstyle[aps,multicol,epsfig]{revtex}
\draft
\begin{document}

\title{Renormalization of quark axial current in the chiral potential model}
\author{X.S. Chen$^{a,c}$, X.B. Chen$^{a,b}$, Amand Faessler$^c$,
        Th. Gutsche$^c$, F. Wang$^a$}
\address{$^a$Department of Physics and Center for Theoretical Physics,
                Nanjing University, Nanjing 210093, China\\
        $^b$Department of Physics, Changsha Institute of Electricity,
                Changsha 410077, China\\
        $^c$Institut f\"ur Theoretische Physik, Universit\"at T\"ubingen,
                Auf der Morgenstelle 14, D-72076 T\"ubingen, Germany}
\date{April 4, 2000}
\maketitle

\begin{abstract}
Non-conserved composite operators like the quark axial current have divergent
matrix elements therefore must be renormalized. We explore how this can be 
done in quark model calculations where the systematic procedure of dimensional 
regularization and minimal subtraction 
is not applicable. We propose a most natural and convenient regularization 
scheme  
of cutting the intermediate quark states over which we sum in loop diagram 
calculations at 
a certain energy. We show that this scheme works perfectly 
for the quark axial current and
we obtain the quark spin contribution to the proton spin: 
$\Delta_u=0.82$, $\Delta_d=-0.43$, $\Delta_s=-0.10$, which is in excellent 
agreement with experiments.
 

\end{abstract}

\begin{multicols}{2}
The expression nucleon spin ``crises'' denotes the findings of  
the European Muon Collaboration (EMC) \cite{EMC} that a small proportion of 
the nucleon spin is carried by the quark spin and that strange quark
polarizes significantly in the nucleon. This has been under hot debate for 
over ten years but one obtained not yet a fully satisfactory description 
(for a review of the nucleon spin problem, see, e.g., \cite{Review}).
It should be emphasized that 
the ``crises'' is not for the fundamental theory of quantum chromodynamics 
(QCD), in the viewpoint of which the nucleon is a complicated object of
quarks and gluons and quarks do not necessarily carry most of the nucleon spin. 
The ``crises'' is, however, for the naive SU(6) quark model which is quite 
successful in many aspects but nevertheless attributes
all the nucleon spin to constituent quarks. 
To explore whether this ``crises'' is {\em real}, 
i.e., whether the SU(6) model could be taken as a good lowest order 
approximation for the nucleon, a natural way is to 
{\em start} from the SU(6) wavefunction and study whether we can 
explain the experimental result of the nucleon spin content by going to 
{\em higher orders}. 

In the past years there has been countless work using quark models 
along this direction (for references see \cite{Review}), 
but the problem is not really resolved. 
The obstacle is that the quark axial current, which is the 
operator for defining
the quark spin contribution to nucleon spin, is a non-conserved composite
operator. Therefore when we go to {\em higher orders} divergent matrix element 
will be encountered and the quark axial current must be renormalized.
But unfortunately, the usual renormalization schemes are not applicable 
in quark model calculations, where instead of divergent loop integrations over
the continuous momentum, we encounter divergent summations over the 
{\em discrete} quark excited states whose wavefunctions are obtain 
{\em numerically}. To explore how one can renormalize a composite operator in 
quark models and what would be the 
result of the {\em renormalized} quark axial charge (i.e., the quark spin 
contribution to nucleon spin) are the aims of this paper.

In the following we will first construct the bare matrix element of the quark
axial current and demonstrate its divergence, then we explore how we can 
renormalize the bare quantity by systematically subtracting a divergent part
and obtain a (finite) physical result.   

The quark spin contribution to the nucleon spin is defined as the quark axial 
charge of the nucleon:
\begin{equation}
\langle ps| \bar{\psi_q} \gamma^\mu \gamma^5 \psi_q|ps\rangle \equiv
\bar u_{ps}\gamma^\mu \gamma^5 u_{ps}\cdot \Delta_q \label{axial},
\end{equation}
where $q=u,d,s$ and $ u_{ps}$ is the nucleon spinor. 
An equivalent but more suitable expression for model calculation is 
\begin{equation}
\Delta_q =\frac{\langle p+| \int d^3 x\bar{\psi_q} \gamma^3 \gamma^5 \psi_q
                |p+\rangle }
         {\langle p+ | p+ \rangle }. \label{Delta_q}
\end{equation}
Here $| p+ \rangle$ is a nucleon state with positive momentum and 
polarization along the third direction. 
We are going to adopt Eq. (\ref{Delta_q}) to 
pursue a perturbative calculation of $\Delta_q$ 
in a chiral potential model. Our model Lagrangian is  
\begin{eqnarray}
{\cal L}&=&\bar{\psi} [ i\partial \hspace*{-2mm}/
                - S(r)- \gamma^0 V(r)]\psi - \nonumber \\
         &&\frac{1}{2F_\pi} \bar{\psi} [ 
        S(r)(\sigma +i\gamma ^5 \lambda^i \phi_i) + 
        (\sigma +i\gamma ^5\lambda^i \phi_i )S(r) ]\psi + \nonumber \\
         &&\frac 12 (\partial_\mu \sigma)^2 + 
           \frac 12 (\partial_\mu \phi_i)^2 -
           \frac 12 m_\sigma^2 \sigma^2 -
           \frac 12 m_i^2\phi_i^2.
            \label{Lagrangian} 
\end{eqnarray}
The model Lagrangian is derived from the $\sigma$ model in which meson fields
are introduced to restore chiral symmetry \cite{Thomas}. The flavor and color
indices for the quark field $\psi$ are suppressed; the scalar term 
$S(r)=cr+m$ represents the linear scalar confinement potential $cr$ and the 
quark mass matrix $m$;
$V(r)=-\alpha /r$ is the Coulomb type vector potential and 
$F_\pi$=93MeV is the pion decay
constant. $\sigma$ and $\phi_i$ ($i$ runs from $1$ to $8$) are the scalar
and pseudoscalar meson fields, respectively and $\lambda_i$ are the Gell-Mann
matrices. The quark-meson interaction term of Eq.(\ref{Lagrangian}) is
symmetrized since the mass matrix $m$ does not commute with all $\lambda_i$ 
for different quark masses.

At zeroth order the nucleon is taken as the usual SU(6) 
three-quark ground state of the Hamiltonian
\begin{equation}
H_q = \int d^3 x \psi ^\dagger  [\vec \alpha \cdot \frac 1i\vec \partial + 
        \beta S(r) +V(r) ] \psi. 
\end{equation}
The diagrams for the numerator and denominator of Eq. (\ref{Delta_q}) up to 
second order are shown in Figs. 1 and 2 respectively. 

\begin{center}
\begin{minipage}{8.5cm}
\begin{figure}
\begin{center}

\psfig{figure=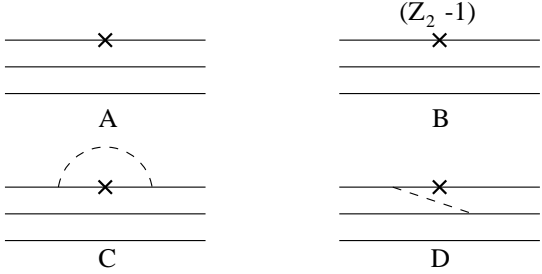,width=8cm}

\bigskip

\begin{tighten}
\caption{Feynman diagrams for the matrix element 
$\langle N| \int d^3 x\bar\psi \gamma^3 \gamma^5 \psi|N\rangle$ 
up to second order; a cross on the quark line denotes the quark axial vertex; 
A is of the zeroth order, B is the renormalization counter term, 
C and D are vertex and exchange diagrams respectively. 
The meson line in C can be a $\pi$, $\eta$, or $\sigma$ (while the
intermediate quark is $u$ or $d$), or a $K$ (while the intermediate quark
is $s$); the meson line in D can only be a $\pi$, $\eta$ or $\sigma $.}
\end{tighten}
\end{center}
\end{figure}
\end{minipage}
\end{center}

\begin{center}
\begin{minipage}{8.5cm}
\begin{figure}
\begin{center}
\psfig{figure=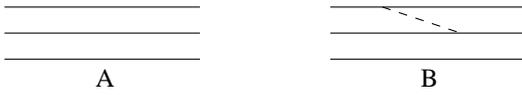,width=7cm}

\bigskip

\begin{tighten}
\caption{Feynman diagrams for the normalization
$\langle N|N\rangle$ up to second order; A is of the zeroth
order which is simply unity, B is the meson exchange
diagram. The meson line in B is a $\pi$, $\eta$, or $\sigma $.}
\end{tighten}
\end{center}
\end{figure}
\end{minipage}
\end{center}

We first discuss how to determine the renormalization constant $Z_2$. The 
mass-shell renormalization scheme is not applicable here, since our 
unperturbed quark basis are confined wavefunctions. 
But we can still use the charge renormalization condition. 
The conserved electromagnetic current of the Lagrangian
of Eq. (\ref{Lagrangian}) is $j^\mu=\sum_q j^\mu_q+j^\mu_\phi$, where $j^\mu_q$
and $j^\mu_\phi$ are the quark and meson current respectively:
\begin{eqnarray}
j^\mu_q   &=& Q_q\bar\psi_q\gamma^\mu\psi_q,  \nonumber \\
j^\mu_\phi&=& e(\phi_1\partial^\mu \phi_2-\phi_2\partial^\mu \phi_1+ 
                \phi_4\partial^\mu \phi_5-\phi_5\partial^\mu \phi_4).
\label{EM}
\end{eqnarray}

The charge renormalization condition is to require that for a quark state
\begin{equation}
\langle q |\int d^3x j^0(x)|q\rangle =Q_q.
\end{equation}
Up to second order this is shown in Fig. 3.  
 
\begin{center}
\begin{minipage}{8.5cm}
\begin{figure}
\begin{center}
\psfig{figure=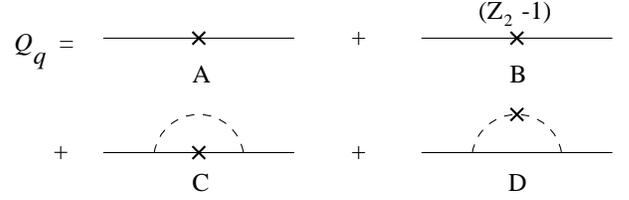,width=8cm}

\bigskip

\begin{tighten}
\caption{Charge renormalization condition; a cross on the quark or meson line 
denotes the zeroth component of the vector current vertex.}  
\end{tighten}
\end{center}
\end{figure}
\end{minipage}
\end{center}
By computing Figs. 3C and 3D
we can determine the renormalization constant $Z_2 $, which is then 
to be used in Fig. 1B.

Next we remark that at zero momentum transfer, the exchange diagram Fig. 1D is 
actually the product of Figs. 1A and 2B, therefore the sum of 
Figs. 1A and 1D over the normalization Figs. 2A and 2B is just Fig. 1A. 
(However this is not true when we calculate the axial form 
factor at finite momentum transfer.) Thus only Fig. 1C is left to  
be evaluated together with Fig. 3C and 3D.

The essential ingredients needed for calculating 
these diagrams are the quark and meson propagators. 
The meson propagator given by the Lagrangian of Eq. (\ref{Lagrangian}) is 
the free propagator:
\begin{equation}
\Delta_{ij}(x_1,x_2) = \frac{i}{(2\pi)^4} \int d^4 q 
                \frac{\delta _{ij} e^{-iq\cdot(x_1-x_2)}}
                {q^2-m_i^2+i\epsilon}.
\end{equation}
Since the non-perturbative confinement is included in
$H_q$ the quark propagator has to be obtained numerically, and in practise we
have to work with time-ordered perturbation theory. We write the solution of
$H_q$ as
\begin{equation}
\psi(x) =\sum _\alpha u_\alpha(x)a_\alpha +
        \sum _\beta v_\beta (x)b^\dagger _\beta, \label{Hq}
\end{equation}
where $u_\alpha(x)=e^{-iE_\alpha t}u_\alpha (\vec x)\tau_\alpha$,
$v_\beta (x)=e^{iE_\beta t}v_\beta (\vec x)\tau_\beta$; $\tau$ is the
flavor wavefunction and $u_\alpha (\vec x) $ and $v_\alpha (\vec x)$
are the spatial wavefunctions. The quark propagator is then
\begin{eqnarray}
D(x_1,x_2)&=& \theta(t_1-t_2)\sum _\alpha u_\alpha(x_1) \bar u_\alpha(x_2)-
        \nonumber \\
        &&  \theta(t_2-t_1)\sum _\beta v_\beta(x_1) \bar v_\beta(x_2).
\end{eqnarray}

It can be shown by carrying out the time and energy integration that 
apart from an isospin factor, Figs. 3C and 3D yield the same expression. 
Therefore we define for Figs. 3C and 3D the pure space-time amplitudes:
\begin{eqnarray}
B_\phi\equiv \frac{1}{F_\pi^2}
       && \int d^3xd^4x_1d^4x_2 \Delta(x_2,x_1) \times \nonumber \\
   &&\bar u_f(x_2) \Gamma_\phi  
         D(x_2,x) \gamma^0 D(x,x_1) 
        \Gamma_\phi u_i(x_1), \label{Fig3}
\end{eqnarray}
where $u_i $ and $u_f $ are the initial and final quark state respectively,
the vertex function $\Gamma_{\pi,K,\eta}=S(r)\gamma^5 $ and
$\Gamma_\sigma=-iS(r)$. All flavor wavefunctions are dropped out and it should
be understood that for $\phi=K$ the intermediate quark is the $s$  
quark and otherwise is the $u $ or $d $ quark.
Accordingly for Fig. 1C we define 
\begin{eqnarray}
A_\phi\equiv \frac{1}{F_\pi^2}
        &&\int d^3xd^4x_1d^4x_2 \Delta(x_2,x_1) \times \nonumber \\
   &&\bar u_f(x_2) \Gamma_\phi  
         D(x_2,x) \gamma^3\gamma^5 D(x,x_1) 
        \Gamma_\phi u_i(x_1). \label{Fig1}
\end{eqnarray}
Now we can express the renormalization constant $Z_2$ and the axial charge 
$\Delta_q$ in terms of $A_\phi$ and $B_\phi$ multiplied by spin and isospin  
factors which are calculated straightforwardly: 
\begin{eqnarray} 
Z_2^{u,d}&=&1-(3B_\pi+2B_K+\frac 13 B_\eta +B_\sigma), \nonumber \\
Z_2^{s}&=&1-(B_K+\frac 43 B_\eta +B_\sigma), \label{Z} \\
\Delta_u&=&\frac{4}{3} f_RZ_2^u
	+\frac 23 A_\pi+\frac 49 A_\eta +\frac 43 A_\sigma 
	\nonumber \\ 
\Delta_d&=&-\frac 13 f_RZ_2^d
        +\frac 73 A_\pi-\frac 19 A_\eta -\frac 13 A_\sigma 
	\nonumber \\ 
\Delta_s&=&2A_K. \label{S}
\end{eqnarray}
Another useful relation is for the nucleon axial charge 
\begin{equation}
g_A=\Delta_u-\Delta_d=\frac 53 f_RZ_2^u
        -\frac 53 A_\pi+\frac 59 A_\eta +\frac 53 A_\sigma \label{gA}
\end{equation} 

In Eqs. (\ref{Z})-(\ref{gA}) 
we have assumed equal masses for $u,d$ and for $\pi^0,\pi^\pm$, therefore  
$Z_2^u =Z_2^d $ which is just a statement of SU(2) symmetry (If SU(3) symmetry
is unbroken $Z_2^s$ would also be the same as $Z_2^{u,d}$); $5/3f_R$ 
is the zeroth order values of $g_A $ and $f_R$ 
is a relativistic reduction factor. 

Since $A_\phi$ and $B_\phi$ correspond to loop diagrams, they would naturally 
be divergent. If the quark axial current {\em were} conserved, 
then the divergence of $A_\phi$ and
$B_\phi$ would automatically cancel in Eq. (\ref{S}) and we get a finite 
result for $\Delta_q$. This is the phenomenon
that conserved operators do not need extra renormalization besides the usual
renormalization of mass, charge and wavefunction. However the quark axial
current is {\em not} conserved, the divergences of $A_\phi $ and
$B_\phi$ will not cancel and the naively obtained $\Delta_q$ in Eq. (\ref{S})  
is divergent, which is just the general case that a composite operator has 
divergent matrix element and needs extra renormalization,
i.e., we are going to subtract a divergent piece from $A_\phi$ and 
$B_\phi$ simultaneously and leave a finite part in Eq.(\ref{S}). This finite 
leftover would depend on how we regularize $A_\phi$ and $B_\phi$ and how much 
is to be subtracted as the divergent part. This is the renormalization 
scheme dependence. 

In the usual plane-wave perturbation theories, we have a most powerful and
systematic renormalization scheme of dimensional regularization and minimal
subtraction (MS) or modified minimal subtraction ($\overline{\rm MS}$), 
but it is evidently not applicable here. We must find a proper 
renormalization scheme for the quark model calculations.
The divergent integration over the (continuous) momentum in the plane-wave 
perturbation theories is here contained in the summation over the (discrete) 
quark intermediate states. Thus we propose to regularize by cutting the 
summation at a certain energy, which is analogous to the lattice regularization 
by a finite lattice spacing. 
In the following we explore how this scheme works in practice.  

Our model parameters are listed in Table I. 
As we demonstrated in a recent paper  \cite{Chen}, 
$A_\phi$ is actually very insensitive to the model 
parameters in the above described energy-cutoff regularization scheme.  
This property is also found for $B_\phi$. Therefore we do not take too much 
effort in choosing the parameters except for a fit to the nucleon and 
$\Delta $ masses with meson exchange potentials.

Figs. 4 and 5 give the numerical results of $B_\phi$ and $A_\phi$ as a
function of the maximum energy up to which one sums the intermediate quark 
states. The divergences are clearly seen. The problem is now how
to determine the cutoff. In general the cutoff might vary from one diagram to 
another, without any guidance we would be at lost.  
Fortunately, besides $\Delta_s=2A_K$, Eq. (\ref{S}) 
provides another clean relation:
\begin{equation}
\Delta_u+4\Delta_d=10A_\pi. \label{clean}
\end{equation}
Thus with the experimental results:
\begin{eqnarray}
\Delta_u&=&0.80(6),~\Delta_d=-0.46(6),~\Delta_s=-0.12(4), \cite{Exp1} 
        \nonumber \\
\Delta_u&=&0.82(6),~\Delta_d=-0.44(6),~\Delta_s=-0.10(4), \cite{Exp2} 
\end{eqnarray}
we can determine the cutoff for $A_K$ and $A_\pi$ using Fig. 5. 
It is amazing to notice that the cutoffs needed for $A_K$ and $A_\pi$ are 
roughly the same, i.e., independent of whether the intermediate quark is 
$u,d$ (for $A_\pi$) or $s$ (for $A_K$) 
and also independent of the meson masses. 
This reminds us that in the MS or $\overline{\rm MS}$ scheme, the
subtraction is independent of the mass parameters. We will therefore choose
a $\phi$-independent cutoff. It is also very interesting to notice that the
cutoff value needed here roughly equals the inverse of the lattice spacing
$a^{-1}$ in lattice QCD calculation of $\Delta_q$. To reduce the number of
parameters, we choose the cutoff for $A_\phi$ 
to be the same as $a^{-1}=1.74$GeV in \cite{Latt}. 
But there remains still one important question 
to be asked: should the cutoff for $A_\phi$ and $B_\phi$ be the same?

\begin{table}
\begin{center}
\begin{tighten}
\caption{Model parameters and basic model predictions; the units for mass,
$\alpha$, and $c$ are MeV, MeV$\cdot$fm, and MeV/fm, respectively; center of
mass corrections are made for the quark core contribution using the
Peierls-Thouless
method [5]. $g_A^{(0)}$ is the zeroth order nucleon axial charge, the
full $g_A$ is obtained later.}
\end{tighten}
\begin{scriptsize}
\begin{tabular}{cccccc}
$m_{u,d}$ &$m_s$  &$m_\pi$ & $m_K $ &$m_\eta$  &$m_\sigma$\\
70 &250&138&495&547&675 \\ \hline
$\alpha$&$c$&$m_N$&$m_\Delta$ &$g_A^{(0)}$ & $g_A$\\
-31.35 &820&939&1232&1.41 & 1.26
\end{tabular}
\end{scriptsize}
\end{center}
\end{table}

It might be taken for granted that they {\em are} the same, (the 
simplest example is that we use the charge renormalization condition to 
determine the renormalization constant and then calculate the matrix element of
the charge operator itself). However, we call special attention here that 
this is {\em not necessarily the case}. The problem is the {\em lack of 
Lorentz covariance} in the model. 

It is actually not difficult to understand this point: If a theory does not 
respect Lorentz covariance, then the renormalization constant determined with
the time and spatial component of the electromagnetic currents are 
possibly {\em different}. Therefore if we calculate the matrix element of the 
charge operator but use the renormalization constant determined through the 
renormalization condition for the spatial component of the vector current, then
we might not obtain the physical charge (in some cases the result may 
still be finite, but this is not enough). 

\begin{center}
\begin{minipage}{8.5cm}
\begin{figure}
\begin{center}
\psfig{figure=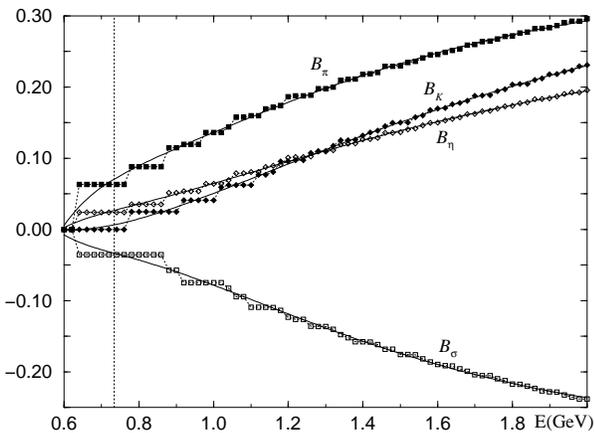,width=8cm}
\bigskip
\begin{tighten}
\caption{Plot of $B_\phi$ as a function of the maximal energy up to which the
intermediate states are summed over.}
\end{tighten}
\end{center}
\end{figure}
\end{minipage}
\end{center}

\begin{center}
\begin{minipage}{8.5cm}
\begin{figure}
\begin{center}
\psfig{figure=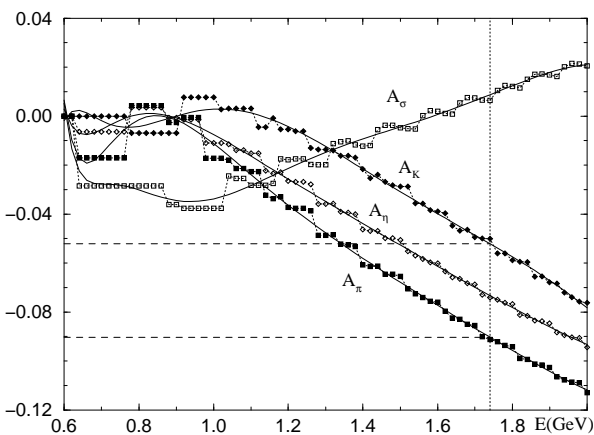,width=8cm}
\bigskip
\begin{tighten}
\caption{Plot of $A_\phi$ as a function of the maximal energy up to which the
intermediate states are summed.}
\end{tighten}
\end{center}
\end{figure}
\end{minipage}
\end{center}

In the present model, Lorentz covariance is violated by the static potentials, 
therefore the renormalization constant determined for the charge might not
be suitable for calculating the axial charge. But since we have presently no 
better method, we must allow different cutoffs for $A_\phi$ and $B_\phi$. 

After describing the formalism, we are now in the position to see
how this renormalization scheme can explain the experimental results. 
As we explained the cutoff for $A_\phi$ is chosen as the inverse 
of the lattice spacing $a^{-1}=1.74$GeV \cite{Latt}. 
Using Eq. (\ref{gA}) and Figs. 4, 5 and requiring $g_A=1.257$, 
the cutoff for $B_\phi$ is determined to be $0.734$GeV. 
Then combining Figs. 4, 5 and using Eq. (\ref{S}), we finally obtain: 
\begin{equation}
\Delta_u=0.823,~\Delta_d=-0.432,~\Delta_s=-0.104.
\end{equation}  

By adjusting only {\em one free parameter} 
(the cutoff for $B_\phi$) to fit $g_A$, we reproduced the 
experimental results of $\Delta_{u,d,s}$ perfectly. 
This can be regarded as a success of our model and renormalization scheme.
Since we have started from the SU(6) wavefunction as the zeroth 
order approximation, we could say that the spin ``crisis'' for the 
naive SU(6) model is {\em not real}.

We end our discussions by emphasizing three points: (1) Non-conserved 
operators have divergent matrix elements and even in model calculations they 
must be consistently renormalized. (2) A natural and convenient 
renormalization scheme in quark model calculations
is to cut the summation over the intermediate quark 
states at a certain energy; a fit to the experimental data reveals that this 
approach shares the same advantage of mass-parameter-independence as in the 
MS or $\overline{\rm MS}$ scheme. (3) However, due to the lack of Lorentz
covariance in quark models, the cutoffs for operators of different Lorentz 
type are not necessarily the same.   

This work is supported by the CNSF (19675018), CSED, CSSTC,
the DFG (FA67/25-1), and the DAAD.

\end{multicols}

\end{document}